\documentclass[12pt]{article}
\usepackage{epsfig}

\def\beq{\begin{equation}}
\def\eeq{\end{equation}}
\def\beqa{\begin{eqnarray}}
\def\eeqa{\end{eqnarray}}

\def\beq{\begin{equation}}
\def\eeq#1{\label{#1}\end{equation}}
\def\eeqn{\end{equation}}

\def\beqa{\begin{eqnarray}}
\def\eeqa#1{\label{#1}\end{eqnarray}}
\def\eeqan{\end{eqnarray}}

\let\bar=\overbar

\def\Dslash{\not{\hbox{\kern-4pt $D$}}}
\def\dslash{\not{\hbox{\kern-2pt $\del$}}}

\def\msb{{\bar{\ssstyle M \kern -1pt S}}}

\def\Title#1{\begin{center} {\Large {\bf #1} } \end{center}}

\begin{document}

\Title{Single top and top pair production}

\bigskip\bigskip

%+\addtocontents{toc}{{\it D. Reggiano}}
%+\label{ReggianoStart}

\begin{raggedright}  

{\it Nikolaos Kidonakis\index{Kidonakis, N.}\\
Kennesaw State University\\
Kennesaw, GA 30144, USA}
\bigskip\bigskip
\end{raggedright}

\section{Introduction}

I present results for single-top and top-pair production at the LHC 
and the Tevatron. Higher-order two-loop corrections are used to achieve 
NNLL resummation, which is then used to derive NNLO soft-gluon corrections. 
Results are presented for total cross sections,  top transverse momentum 
distributions, and top rapidity distributions. All results are in 
excellent agreement with data from the LHC and the Tevatron. I also clarify 
the differences between various methods in top-pair production and their 
relation to exact NNLO results.

The LO partonic processes for single-top production are 
$qb \rightarrow q' t$ and ${\bar q} b \rightarrow {\bar q}' t$, i.e. the 
$t$ channel, dominant at the Tevatron and the LHC; 
$q{\bar q}' \rightarrow {\bar b} t$, i.e. the $s$ channel, numerically 
small at the 
Tevatron and the LHC; and associated $tW$ production, $bg \rightarrow tW^-$,
very small at the Tevatron but significant at the LHC.
The LO partonic processes for 
top-pair production are $q{\bar q} \rightarrow t {\bar t}$, 
dominant at the Tevatron; and $gg \rightarrow t {\bar t}$, dominant at the LHC.

QCD corrections are significant for single top and top pair production.
Soft-gluon corrections from emission of soft (low-energy) gluons
appear as $\ln^k(s_4/m^2)/s_4$,  
with $k \le 2n-1$ and $s_4$ the distance from threshold, 
and are dominant near threshold.
We resum these soft corrections using factorization and renormalization-group 
evolution. Complete results are available at NNLL using two-loop soft 
anomalous dimensions for both top-pair \cite{NKtt,NKy} and 
single-top production \cite{NK2ls,NK2lw,NK2lt}.
Approximate NNLO differential cross sections are derived from the 
expansion of the resummed cross section; 
see Ref. \cite{NK12} for the newest results.
This is the only calculation using the standard moment-space resummation 
in pQCD for partonic threshold at the double-differential cross section level.
The soft-gluon approximation works very well for both Tevatron 
and LHC energies; there is less than 1\% difference between NLO approximate 
and exact cross sections, and similarly for differential distributions \cite{NK12}.

\let\thefootnote\relax\footnote{Proceedings of CKM 2012,
the 7th International Workshop on the CKM Unitarity Triangle, 
Cincinnati, Ohio, USA, 28 September - 2 October 2012}

\section{Single top production}

\begin{figure}[htb]
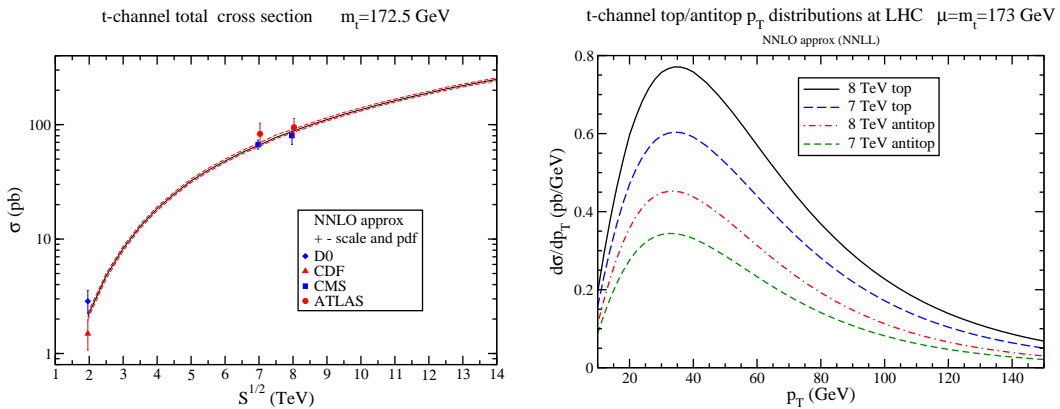

\begin{center}
\epsfig{file=tchtotalSlhcplot.eps,height=2.1in}
\hspace{3mm}
\epsfig{file=pttchtopatoplhcplot.eps,height=2.1in}
\caption{Single-top $t$-channel cross sections (left) and $p_T$ distributions (right).}
\label{tchannel}
\end{center}
\end{figure}

We begin with the $t$ channel. On the left plot of Fig. \ref{tchannel} we 
show the total approximate NNLO $t$-channel cross section versus  
collider energy with $m_t=172.5$ GeV and MSTW2008 pdf \cite{MSTW} and compare it with recent 
Tevatron \cite{CDFts,D0ts} and LHC \cite{ATLAStch,CMStch} 
data. The agreement between theory and 
experiment is very good. The $t$-channel top and antitop $p_T$ distributions 
at the LHC are shown in the right plot.

\begin{table}[b]
\begin{center}
\begin{tabular}{c|c|c|c}
LHC  & $t$ &  ${\bar t}$ & Total (pb) \\ 
\hline
7 TeV  & $43.0 {}^{+1.6}_{-0.2} \pm 0.8$ 
& $22.9 \pm 0.5 {}^{+0.7}_{-0.9}$
& $65.9 {}^{+2.1}_{-0.7} {}^{+1.5}_{-1.7}$
\\
8 TeV  & $56.4 {}^{+2.1}_{-0.3} \pm 1.1$ 
& $30.7 \pm 0.7 {}^{+0.9}_{-1.1}$
& $87.2 {}^{+2.8}_{-1.0} {}^{+2.0}_{-2.2}$
\end{tabular}
\caption{$t$-channel cross sections $\pm$ scale $\pm$ pdf errors
for $m_t=173$ GeV with MSTW2008 NNLO 90\% CL pdf \cite{MSTW}.}
\end{center}
\end{table}

Table 1 shows the separate single-top and single-antitop $t$-channel cross 
sections and their sum at 7 and 8 TeV LHC energy for $m_t=173$ GeV.
The ratio $\sigma(t)/\sigma({\bar t})= 1.88{}^{+0.11}_{-0.09}$ at 7 TeV,  
which compares well with the ATLAS result  $1.81{}^{+0.23}_{-0.22}$ 
\cite{ATLASratio}. 

For the $s$-channel, the total cross sections 
(single top+antitop) at the LHC with $m_t=173$ GeV are 
$4.56 \pm 0.07 {}^{+0.18}_{-0.17}$ pb at 7 TeV  
and $5.55 \pm 0.08 \pm 0.21$ pb at 8 TeV.

\begin{figure}[htb]
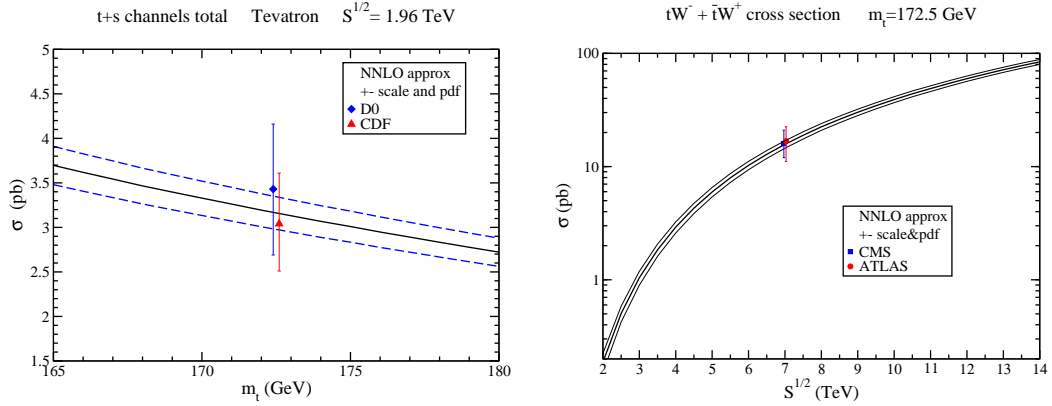

\begin{center}
\epsfig{file=tchschtevplot.eps,height=2.1in}
\hspace{3mm}
\epsfig{file=tWtotalSlhcplot.eps,height=2.1in}
\caption{Single-top $t$- plus $s$-channel (left) and $tW$ (right) cross sections.}
\label{tsWchannels}
\end{center}
\end{figure}

At the Tevatron at 1.96 TeV for $m_t=173$ GeV, the $t$-channel total 
(single top and single antitop) cross section is  
$2.08 {}^{+0.00}_{-0.04} \pm 0.12$ pb;  
the $s$-channel total is $1.05 {}^{+0.00}_{-0.01} \pm 0.06$ pb; and the 
sum of the two channels is $3.13 {}^{+0.00}_{-0.05} \pm 0.18$ pb. 
The sum of the $t$- and $s$-channel cross sections at the Tevatron 
is plotted in the left plot of Fig. \ref{tsWchannels} as a function of $m_t$ 
and compared with CDF \cite{CDFts} and D0 \cite{D0ts} data.

For associated $tW^-$ production at the LHC with $m_t=173$ GeV, the 
cross section at 7 TeV is $7.8 \pm 0.2 {}^{+0.5}_{-0.6}$ pb 
and at 8 TeV it is $11.1 \pm 0.3 \pm 0.7$ pb. 
The cross section for ${\bar t}W^+$ production is identical.
In the right plot of Fig. \ref{tsWchannels} we display the total 
$tW^-$ plus ${\bar t}W^+$ cross section with $m_t=172.5$ GeV versus LHC energy 
and compare it with recent ATLAS \cite{ATLAStW} 
and CMS \cite{CMStW} results - the agreement is excellent.

\section{Top pair production}

\begin{figure}[htb]
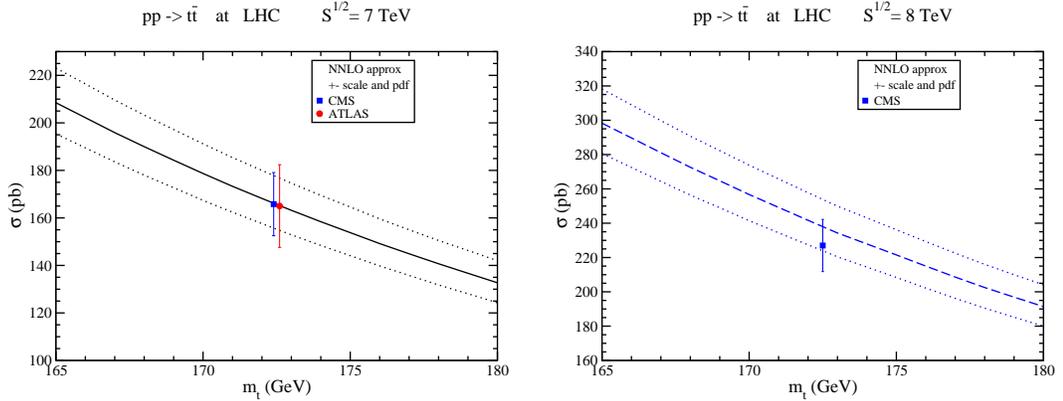

\begin{center}
\epsfig{file=ttb7lhcplot.eps,height=2.1in}
\hspace{3mm}
\epsfig{file=ttb8lhcplot.eps,height=2.1in}
\caption{Top pair cross sections at the LHC at 7 TeV (left) and 8 TeV (right).}
\label{ttblhc}
\end{center}
\end{figure}

The total $t{\bar t}$ cross section with $m_t=173$ GeV
at the LHC is  $163 {}^{+7}_{-5} \pm 9$ pb at 7 TeV; 
and $234 {}^{+10}_{-7} \pm 12$ pb at 8 TeV.
Figure 3 shows the cross section as a function of $m_t$ 
at 7 TeV (left) and 8 TeV (right) together with LHC data 
\cite{ATLASttb,CMSttb}.
The agreement of theory with experiment is very good and the 
uncertainties are similar.

\begin{figure}[htb]
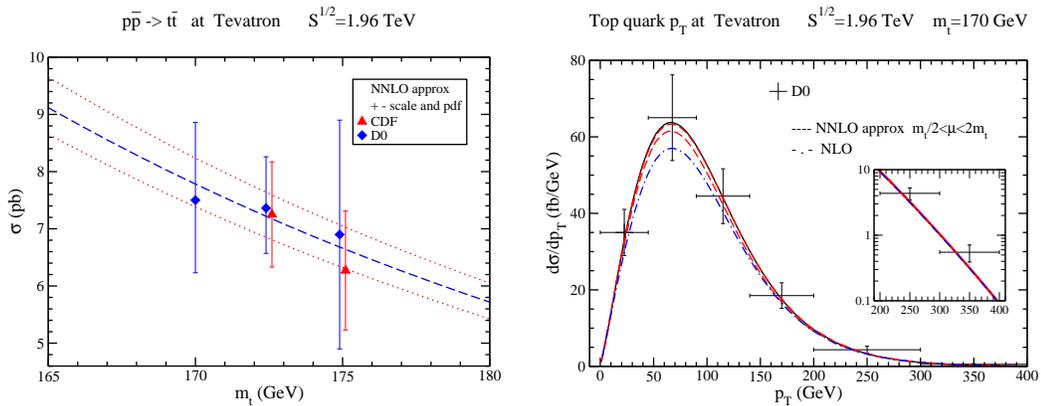

\begin{center}
\epsfig{file=ttbtevplot.eps,height=2.1in}
\hspace{3mm}
\epsfig{file=ptD0tevnewplot.eps,height=2.1in}
\caption{Top pair cross sections (left) and top $p_T$ distributions (right) 
at Tevatron.}
\label{ttbtev}
\end{center}
\end{figure}

The $t{\bar t}$ cross section at the Tevatron at 1.96 TeV  
is $7.08 {}^{+0.00}_{-0.24} {}^{+0.36}_{-0.27}$ pb for $m_t=173$ GeV. 
The left plot of Fig. 4 shows the cross section and compares it to Tevatron 
data \cite{CDFttb,D0ttb}. 
The right plot shows the top $p_T$ distribution together with D0 data 
\cite{D0pt}. Again the agreement of theory and experiment is excellent.

The resummation approaches that have appeared in \cite{NKtt,HATHOR,AFNPY,BFKS}
differ from each other in many respects. Differences include the use  
of moment-space pQCD \cite{NKtt,HATHOR} versus SCET \cite{AFNPY,BFKS}; 
doing the resummation for the double-differential cross section 
\cite{NKtt,AFNPY} versus for the total cross section only \cite{HATHOR,BFKS}; 
inclusion of subleading terms, etc. More discussion can be found in 
\cite{NK12,NKBP}. 
All results presented in this paper are approximate NNLO 
from moment-space NNLL resummation 
for the double-differential cross section in single-particle-inclusive (1PI) 
kinematics.
 
The result of Ref. \cite{NKtt} is very close to the partially exact NNLO 
(exact for $q{\bar q}$ plus approximate for $gg$) of \cite{BCM} at Tevatron energy: 
$7.08$ vs $7.07$ pb, with similar scale uncertainty.   
It is claimed in \cite{BCM} that the threshold approximation is not very 
good; however that only applies to the resummation method used in \cite{BCM}.
As discussed in detail in \cite{NK12} and in \cite{NKBP}, the various  
resummation formalisms are both numerically and theoretically very different. 
Thus, the claim in \cite{BCM} does not necessarily apply to other methods, 
and it is most certainly irrelevant to the method used here; as shown 
explicitly in \cite{NK12} and previous work, the method that we use here 
provides an excellent approximation (less that 1\% difference between exact 
and approximate results both at NLO and at NNLO), 
which is significantly better than in \cite{BCM}. 
This was expected from the study of 1PI and PIM results in \cite{NKRV1}
(see also discussion in \cite{NKtt,NK12,NKRV2}). 
A double-differential calculation 
for partonic (not absolute) threshold as used here and in \cite{NKtt,NKy,NK12} 
has a lot more theoretical/analytical information (also useful for deriving 
distributions)  
and potential for numerical accuracy than one for the total cross section 
only as used in \cite{BCM}. 
This is a point often misunderstood and not emphasized in the literature.
We also note that once NNLO is fully known, the next step will be to add 
the approximate NNNLO corrections (see \cite{NKNNNLO} for some early NNNLO 
results).

\begin{figure}[htb]
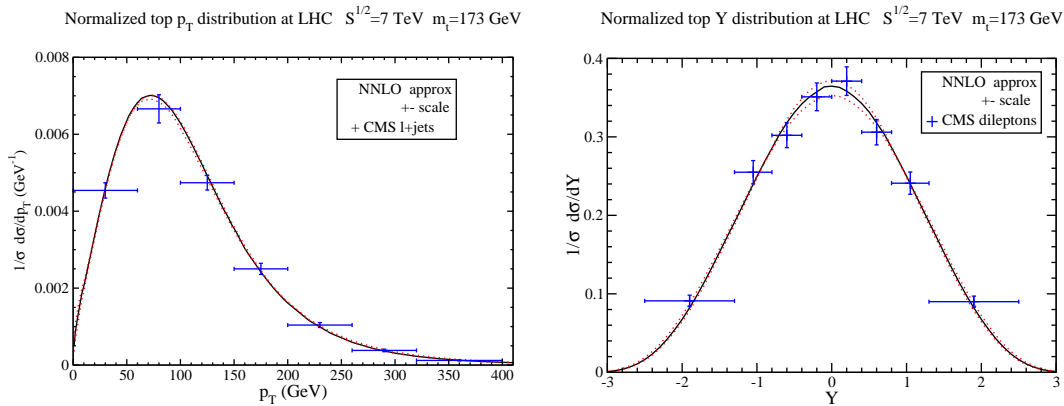

\begin{center}
\epsfig{file=pt7lhcnormCMSljetplot.eps,height=2.1in} 
\hspace{3mm}
\epsfig{file=y7lhcnormCMSdileptplot.eps,height=2.1in}
\caption{Top-quark $p_T$ (left) and rapidity (right) distributions at the LHC.}
\label{pty}
\end{center}
\end{figure}

The normalized top quark $p_T$ distribution at the LHC is shown in the 
left plot of Fig. 5 while the normalized top quark rapidity distribution 
is shown on the right. Both distributions predict very well the CMS 
results \cite{CMSpty}, also shown on the plots.

\bigskip
This material is based upon work supported by the National Science Foundation 
under Grant No. PHY 1212472.

\end{document}